# Privacy Threats on the Internet of Medical Things

Nyteisha Bookert, *North Carolina Agricultural and Technical State University*
Mohd Anwar, *North Carolina Agricultural and Technical State University*

## Abstract

The Internet of Medical Things (IoMT) is a frequent target of attacks -- compromising both patient data and healthcare infrastructure. While privacy-enhanced technologies and services (PETS) are developed to mitigate traditional privacy concerns, they cannot be applied without identifying specific threat models. Therefore, our position is that the new threat landscape created by the relatively new and underexplored IoMT domain must be studied. We briefly discuss specific privacy threats and threat actors in IoMT. Furthermore, we argue that the privacy policy gap needs to be identified for the IoMT threat landscape.

## 1. Introduction

The Internet of Medical Things (IoMT), IoT devices for healthcare, is a frequent target of attacks by hackers, compromising both patient data and patient care (healthcare infrastructure). In 2019, 41.3 million healthcare records were breached [1]. Several companies have recalled medical devices due to security and privacy issues [2]. Existing work has suggested several solutions to address privacy concerns, including encryption [3]–[5], authentication [6]–[8], access control [9]–[11], and de-identification [12]–[14]. While efforts are being made to mitigate general privacy concerns such as confidentiality and control of information, we must evaluate the new threat landscape created by the relatively new and underexplored IoMT domain. We discuss privacy threats and threat actors specific to IoMT. We argue that the unique challenges and privacy threats specific to the IoMT domain should be studied to ensure appropriate solutions. This paper first provides an overview of the IoMT architecture, then discusses privacy threats specific to IoMT, and finally addresses potential threat actors in IoMT.

## 2. Internet of Medical Things Architecture

We consider a five-layer IoT architecture of IoMT [15], [16]. The layers are perception, network, middleware, application, and business. The perception layer represents the physical layers. It contains sensors and actuators. It is responsible for collecting data and sending it to the network layer. The network layer is responsible for transferring the data from the perception layer to the middleware layer. It implements network protocols such as RFID, Wi-Fi, and Bluetooth. The middleware receives the data from the perception layer. It processes the data and provides access control. It delivers the service. The application layer is where the user interacts with and queries the data. This layer describes the various use cases for the data. The business layer is responsible for extracting knowledge from the data. The decision-making and business modeling take place in this layer.

## 3. Privacy Threats in IoMT

We will briefly discuss privacy threats and how they apply to IoMT. We adapt two threat methodologies, LINDDUNN [17] and Ziegeldorf et al. [18], to discuss privacy threats in IoMT devices. The description of each threat and how it applies to IoMT are as follows.

- *Identification* refers to the ability to identify an entity or person in a system or data. The usage of personally identifiable information (PII) and other identifiers increases the likelihood of identification in IoMT. Additionally, the automatic collection of data leads to an increase in the volume and variety of data. The diversity in data introduces the possibility of new quasi-identifiers. In IoMT, identifiers exist in the physical layer (e.g., mac address, device identifier), network layer (e.g., Wi-Fi access points, IP address), and application layer (e.g., user's email address, user's phone number). Additionally, the data processed in the middleware and business layer may contain quasi-identifiers (e.g., a newborn's weight, mother's age, and hospital [19].

- *Linkage* refers to being able to link or map an entity or a person from independent datasets. Users may interact with a variety of services assuming the data is separate. However, an entity may combine the data and reveal links between the data. The data collected by IoMT devices are diverse and vast. When combined with existing data, new privacy threats emerge. For example, Alphabet owns Fitbit and Google. The data generated by the fitness tracker could be linked to the data from a user's internet search on Google using the user's

Mobile phone equipment ID, Android ID, or MAC address.

- *Disclosure* of information is the ability of an unauthorized party or system to learn the contents or information about a person or entity. Some examples of the disclosure include a nurse accessing the records of a patient not under their charge, an adversary gaining access to pharmaceutical research data, and a company selling data without the consumers' permission. Additionally, IoT devices may contain privacy-violating interactions and presentations that reveal information to others. Medical devices and applications transition through various stages such as borrowing, selling, and destroying. If handled incorrectly, the data is at risk of being disclosed to unauthorized individuals.

- *Localization and Tracking* is the ability to track individuals through time and space. Devices and applications can capture GPS, Cell Towers, and Wi-Fi locations allowing organizations access to individuals' movements. This threat exists for wearables and implantables.

- *Profiling* is creating a database of information on an individual. The previous privacy threats (i.e., identification, linkage, localization and tracking) make profiling possible. For example, tracking an individual's movement can provide a detailed history of a person's habits and routines. Profiling can lead to discrimination and unsolicited marketing. The privacy threats at this stage include aggregation, profiling, data linking, and data sharing.

- *Non-repudiation* is a security property that enables the ability to verify an action within a system. However, it creates a privacy threat to a user's plausible deniability. Consider a patient with a reoccurring appointment with a specialized doctor for a rare disorder. If an attacker can prove that the patient is meeting the doctor, they can learn the patient's medical condition. A user should have the capacity to deny the claims of an attacker.

- *Detectability/Inventory Attacks* refer to discovering whether a device or application is in an environment.

- *Unawareness* refers to a person not knowing about collection practices, purposes, and sharing practices. There are a variety of sensors in IoMT devices that could collect data outside the scope of their intended functionality and service. Additionally, users may not know how the organization uses the data or whom it is shared.

- *Non-compliance* refers to a system that does not meet the requirements of applicable regulations and laws. IoMT devices must abide by various health laws, technology laws, federal, state, and local laws depending on where they are owned and operated. Due to the complexity and evolving legal requirements, devices may not adhere to all legal requirements.

## 4. IoMT Threat Actors

We identify threat actors in the IoMT realm. This is important because IoMT developers and manufacturers must consider who may infringe upon privacy. Threat actors are as follows:

- Adversarial: The actor does not know the individual and is acting with malicious intent. Three types of adversarial intruders exist: individual, organized group, and state-sponsored [20]. An organized group may seek the medical records from a hospital to sell on the black market such as the dark web.

- Abusers: The actor knows the individual and is acting to create fear or control. For example, "an authenticated but adversarial user of a victim's device or account who carries out attacks by interacting with the standard user interface" [21]. Abusers may have access to the IoMT device or use that access to physically prevent the use of the device or companion mobile application.

- Organizations: The actor works for their benefit and may not intentionally seek to infringe upon privacy. Organizations may collect data to gain insight into users' habits to improve devices or conduct research. However, the data collected could allow for tracking and profiling.

## 5. Summary

Privacy threat models and threat actors for IoMT are explored in this paper. It is important to consider societal and policy threats along with technical threats. We must also consider how the threats evolve in a particular domain, and threat models must be flexible and extendable.

## Acknowledgements

This work was supported partially by the U. S. Department of Education under the Title III Historically Black Graduate Institutions (HBGI) grant. The views and conclusions contained herein are those of the authors and should not be interpreted as necessarily representing the official policies or endorsements, either expressed or implied, of the U. S. Government. The U. S. Government is authorized to reproduce and distribute reprints for governmental purposes notwithstanding any copyright annotation therein.